# Thermodynamic theory of epitaxial ferroelectric thin films with dense domain structures

V. G. Koukhar,[1] N. A. Pertsev,[1,2,*] and R. Waser[2,3]

[1]*A.F. Ioffe Physico-Technical Institute, Russian Academy of Sciences, 194021 St. Petersburg, Russia*
[2]*Institut für Werkstoffe der Elektrotechnik, RWTH Aachen University of Technology, D-52056 Aachen, Germany*
[3]*Elektrokeramische Materialen, Institut für Festkörperforschung, Forschungszentrum Jülich, D-52425 Jülich, Germany*

A Landau-Ginsburg-Devonshire-type nonlinear phenomenological theory is presented, which enables the thermodynamic description of dense laminar polydomain states in epitaxial ferroelectric thin films. The theory explicitly takes into account the mechanical substrate effect on the polarizations and lattice strains in dissimilar elastic domains (twins). Numerical calculations are performed for $PbTiO_3$ and $BaTiO_3$ films grown on (001)-oriented cubic substrates. The "misfit strain-temperature" phase diagrams are developed for these films, showing stability ranges of various possible polydomain and single-domain states. Three types of polarization instabilities are revealed for polydomain epitaxial ferroelectric films, which may lead to the formation of new polydomain states forbidden in bulk crystals. The total dielectric and piezoelectric small-signal responses of polydomain films are calculated, resulting from both the volume and domain-wall contributions. For $BaTiO_3$ films, strong dielectric anomalies are predicted at room temperature near special values of the misfit strain.

## I. INTRODUCTION

In thin films of perovskite ferroelectrics epitaxially grown on many substrates, the lattice misfit of the epitaxial couple creates a driving force for the formation of regular ferroelastic domain (twin) structures below the phase transition temperature. The twinning of epitaxial layers was predicted theoretically by Roitburd already in 1976,[1] and during the past decade domain patterns were found in $PbTiO_3$, $Pb(Zr_xTi_{1-x})O_3$, $(Pb_{1-x}La_x)TiO_3$, $BaTiO_3$, $KNbO_3$, and $SrBi_2Ta_2O_9$ films grown on various single-crystalline substrates.[2-20]

The experimental observations triggered intensive theoretical studies of the statics and dynamics of elastic domains (twins) in epitaxial ferroelectric and ferroelastic thin films.[1,21-37] In the general case, the problem is very complicated because of substantial inhomogeneity of internal mechanical stresses in polydomain (twinned) films. For a domain configuration with an arbitrary geometry, the solution may be found with the aid of the dislocation-disclination modeling of the stress sources existing in epitaxial layers at the film/substrate interface and the junctions of ferroelastic domain walls.[25,26,30,37] The complexity of this theoretical approach, however, reduces to a reasonable level only in a linear elastic approximation, which neglects deviations of the order parameters in thin films from their equilibrium values in stress-free bulk crystals. In the case of ferroelectric films, on the contrary, the mechanical substrate effect may strongly change the polarization components, as shown recently for single-domain films with the aid of a nonlinear thermodynamic theory.[38,39] Therefore, a rigorous theoretical analysis, which does not employ the linear approximation, is required to describe polydomain states in epitaxial ferroelectric films correctly.

The basis for this analysis is provided by the Landau-Ginsburg-Devonshire phenomenological theory, which was widely used in the past to explain physical properties of bulk ferroelectrics.[40-42] To make the nonlinear description of polydomain films mathematically feasible, it is necessary to assume the polarization and strain fields to be homogeneous within each domain. This approximation may be justified for dense laminar polydomain states, where the domain widths are much smaller than the film thickness.[36] Such domain structures become energetically most favorable in epitaxial films with a thickness larger than about 100 nm.[26]

In this paper, a nonlinear Landau-Ginsburg-Devonshire-type thermodynamic theory is developed for polydomain epitaxial films of perovskite ferroelectrics. The method of theoretical calculations is reported, which makes it possible to determine polarizations, lattice strains, and mechanical stresses inside dissimilar domains forming dense laminar structures (Sec. II). For $PbTiO_3$ and $BaTiO_3$ films grown on dissimilar cubic substrates, the "misfit strain-temperature" phase diagrams are constructed, which show the stability ranges of various possible polydomain and single-domain states (Sec. III). The small-signal dielectric responses of polydomain $PbTiO_3$ and $BaTiO_3$ epitaxial films are calculated numerically, and their changes at the misfit-strain-driven structural transformations are discussed (Sec. IV). The piezoelectric properties of polydomain ferroelectric thin films are also considered (Sec. V). Finally, some general features of the polydomain (twinned) states in epitaxial ferroelectric films are discussed, and the theoretical predictions are compared with available experimental data (Sec. VI).

It should be emphasized that the developed nonlinear theory enables the calculation of the total dielectric and piezoelectric responses of polydomain thin films with the account of the mechanical substrate effect. This feature of the theory demonstrates its great practical importance, because ferroelectric thin films have many possible applications in advanced microelectronic and micromechanical devices.[43,44]

## II. METHOD OF THEORETICAL CALCULATIONS

We will discuss single crystalline epitaxial films grown in a paraelectric state on much thicker dissimilar substrates. During the cooling from the deposition temperature $T_g$, the paraelectric to ferroelectric phase transition is assumed to take place in the epitaxial layer. This transition leads to the formation of either a single-domain or a polydomain state. In the simplest case, the latter is composed of alternating domains of two different types. According to the available experimental data, the ferroelastic domain structure of an epitaxial film may be modeled by a periodic array of parallel flat domain boundaries. The geometry of this laminar pattern is defined by the domain-wall periodicity $D$ and the volume fraction $\phi$ of domains of the first type in the film.

In relatively thick films the domain widths are expected to be much smaller than the film thickness $H$.[23,26] In this case of a dense structure ($D << H$), the polarization and strain fields become almost uniform within each domain in the inner region of the film, because highly inhomogeneous internal fields can exist only in two thin layers ($h \sim D$) near the film surfaces (see Fig. 1a).[1] Therefore, the distribution of the energy density in a polydomain film may be regarded as *piecewise homogeneous* when calculating the total free energy $\Im$ of the epitaxial system. Indeed, the contribution of the inner region to the energy $\Im$ is about $H/D$ times larger than the contribution of the surface layers so that the latter may be neglected at $D << H$ in the first approximation. The elastic energy stored in a substrate having elastic compliances similar to those of the film can be ignored on the same grounds. The contribution of the self-energies of domain walls also can be neglected in the range of film thicknesses $H >> 100\ H_0$, where the condition $D << H$ becomes valid.[26] This contribution scales as $\sqrt{H}$ here because it is proportional to $H/D$ and the equilibrium domain period $D \sim \sqrt{H}$.[26] Accordingly, the film elastic strain energy is $\sqrt{H/H_0}$ times larger than the overall self-energy of domain walls, where $H_0 \sim 1$ nm is the characteristic film thickness described in Ref. 26.

In the resulting approximation, equilibrium values of polarization components $P_i$ ($i$ = 1,2,3) and lattice strains $S_n$ ($n$ =1,2,3,...,6 in the Voigt matrix notation) inside domains of two types and their equilibrium volume fractions become independent of the domain period $D$ and film thickness $H$. The calculation of these parameters defining piecewise homogeneous fields in the inner region of a polydomain film represents the goal of the present theory. Determination of the equilibrium domain period $D^*$ remains beyond the scope of this theory because $D^*$ is governed by the competition of the overall self-energy of domain walls and the energy stored in the surface layers. However, the equilibrium period $D^*$ may be evaluated in the linear elastic approximation. The calculations show[26] that the inequality $D^* << H$ indeed holds for sufficiently thick films ($H$ >> 100 nm in the case of $PbTiO_3$ and $BaTiO_3$ films).

For the correct thermodynamic description of polydomain thin films, an appropriate form of the free-energy function must be chosen, which corresponds to the actual mechanical and electric boundary conditions of the problem. In this paper, we shall assume that the film is kept under an external electric field $\mathbf{E}_0$, but the film/substrate system is not subjected to external mechanical forces. Since the work done by extraneous mechanical sources equals zero, the free energy $\Im$ of the heterostructure can be derived solely from the distribution of the Helmholtz free-energy density $F$ in its volume. This density, however, must be taken in a modified form $\tilde{F} = F - \sum_{i=1}^{3} E_i(\varepsilon_0 E_i + P_i)$,[45] where $\mathbf{E}$ is the *internal* electric field and $\varepsilon_0$ is the permittivity of the vacuum, because in our case electrostatic potentials of the electrodes are kept fixed but not their charges.

Since we are considering here only periodic domain structures, the minimization of the total free energy $\Im$ may be replaced by the minimization of the energy density $<\tilde{F}>$ averaged over the domain period $D$ and the film thickness $H$. For dense structures, the mean density $<\tilde{F}>$ can be evaluated from the relation

$$<\tilde{F}> = \phi\,\tilde{F}' + (1-\phi)\,\tilde{F}'', \qquad (1)$$

where $\tilde{F}'$ and $\tilde{F}''$ are the characteristic energy densities in the inner regions of domains of the first and second type, within which the polarization and strain fields are almost uniform. (The primed and double-primed quantities below refer to these two types of domains in the same sense.)

We focus now on (001)-oriented perovskite films epitaxially grown on a cubic substrate with the surface parallel to the (001) crystallographic planes. Then the Helmholtz free-energy function $F$ may be approximated by a six-degree polynomial in polarization components $P_i$.[40] Since for perovskite ferroelectrics the Gibbs energy function $G$ is defined better than $F$,[41,42] it is convenient to derive the Helmholtz free energy $F$ via the inverse Legendre transformation of $G$. This procedure gives $F = G + \sum_{n=1}^{6} \sigma_n S_n$ so that for ferroelectrics with a cubic paraelectric phase we obtain the energy density $\tilde{F}$ as

$$\begin{aligned}\tilde{F} &= \alpha_1(P_1^2 + P_2^2 + P_3^2) \\ &+ \alpha_{11}(P_1^4 + P_2^4 + P_3^4) \\ &+ \alpha_{12}(P_1^2P_2^2 + P_1^2P_3^2 + P_2^2P_3^2) \\ &+ \alpha_{111}(P_1^6 + P_2^6 + P_3^6) \\ &+ \alpha_{112}[P_1^4(P_2^2 + P_3^2) + P_2^4(P_1^2 + P_3^2) + P_3^4(P_1^2 + P_2^2)] \\ &+ \alpha_{123}P_1^2P_2^2P_3^2 \\ &+ \frac{1}{2}s_{11}(\sigma_1^2 + \sigma_2^2 + \sigma_3^2) + s_{12}(\sigma_1\sigma_2 + \sigma_1\sigma_3 + \sigma_2\sigma_3) \\ &+ \frac{1}{2}s_{44}(\sigma_4^2 + \sigma_5^2 + \sigma_6^2) \\ &- \frac{1}{2}\varepsilon_0(E_1^2 + E_2^2 + E_3^2) - E_1P_1 - E_2P_2 - E_3P_3, \qquad (2)\end{aligned}$$

where $\sigma_n$ are the internal mechanical stresses in the film, $\alpha_1$, $\alpha_{ij}$, and $\alpha_{ijk}$ are the dielectric stiffness and higher-order stiffness coefficients at constant stress,[41,42] and $s_{mn}$ are the elastic compliances at constant polarization. The dielectric stiffness $\alpha_1$ should be given a linear temperature dependence $\alpha_1 = (T - \theta)/2\varepsilon_0 C$ based on the Curie-Weiss law ($\theta$ and $C$ are the Curie-Weiss temperature and constant). It should be noted that Eq. (2) is written in the crystallographic reference frame ($x_1$, $x_2$, $x_3$) of the paraelectric phase, and we shall take the $x_3$ axis to be orthogonal to the substrate surface below (Fig. 1).

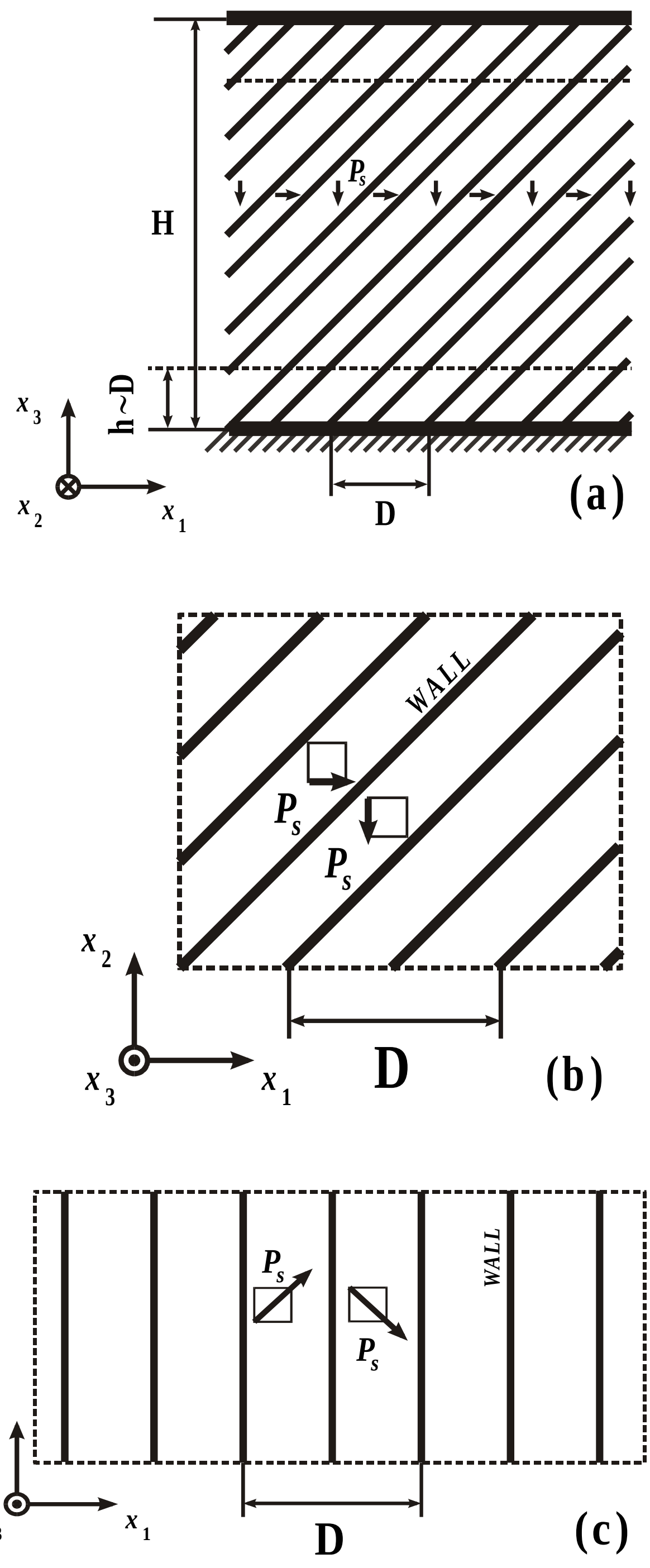

**FIG. 1.** Schematic drawing of the dense $c/a/c/a$ (a), $a_1/a_2/a_1/a_2$ (b), and $aa_1/aa_2/aa_1/aa_2$ (c) domain structures in epitaxial thin films of perovskite ferroelectrics. $H$ is the film thickness, $D$ << $H$ is the domain-wall periodicity. The dashed line indicates a thin layer near the film/substrate interface, where internal stresses are highly inhomogeneous. The $x_3$ axis of the rectangular reference frame ($x_1$, $x_2$, $x_3$) is orthogonal to the interface, whereas the $x_1$ and $x_2$ axes are parallel to the in-plane crystallographic axes of the film prototypic cubic phase.

Using the mechanical and electric boundary conditions of the problem, it is possible to eliminate the stresses $\sigma_i', \sigma_i''$ and internal electric fields $E_i', E_i''$ from the final expression for the mean energy density $< \tilde{F} >$ in the film, which follows from Eqs. (1)-(2). Since changes of the in-plane sizes and shape of the film during its cooling from the deposition temperature are controlled by a much thicker substrate, the mean in-plane film strains $<S_1>$, $<S_2>$, and $<S_6>$ must be fixed quantities at a given temperature. For films grown on (001)-oriented cubic substrates, this strain condition gives

$$\phi S_1' + (1-\phi)S_1'' = S_m, \quad \phi S_2' + (1-\phi)S_2'' = S_m,$$
$$\phi S_6' + (1-\phi)S_6'' = 0, \qquad (3)$$

where $S_m = (b^* - a_0)/b^*$ is the misfit strain[38,39] in the heterostructure ($b^*$ is the substrate effective lattice parameter allowing for the possible presence of misfit dislocations at the interface,[24] and $a_0$ is the equivalent cubic cell constant of the free standing film). Using the thermodynamic relations $S_n = - \partial G/\partial \sigma_n$,[42] we can express the lattice strains through polarization components and mechanical stresses:

$$S_1 = s_{11}\sigma_1 + s_{12}(\sigma_2 + \sigma_3) + Q_{11}P_1^2 + Q_{12}(P_2^2 + P_3^2), \qquad (4)$$

$$S_2 = s_{11}\sigma_2 + s_{12}(\sigma_1 + \sigma_3) + Q_{11}P_2^2 + Q_{12}(P_1^2 + P_3^2), \qquad (5)$$

$$S_3 = s_{11}\sigma_3 + s_{12}(\sigma_1 + \sigma_2) + Q_{11}P_3^2 + Q_{12}(P_1^2 + P_2^2), \qquad (6)$$

$$S_4 = s_{44}\sigma_4 + Q_{44}P_2P_3, \qquad (7)$$
$$S_5 = s_{44}\sigma_5 + Q_{44}P_1P_3, \qquad (8)$$
$$S_6 = s_{44}\sigma_6 + Q_{44}P_1P_2, \qquad (9)$$

where $Q_{ij}$ are the electrostrictive constants of the paraelectric phase. Substituting Eqs. (4), (5), and (6) into Eqs. (3), we obtain the first three conditions imposed on the mechanical stresses $\sigma_i', \sigma_i''$ inside domains of the first and second type. Another three conditions follow from the absence of tractions on the free surface of the film in our case. The mean values of the stress components $\sigma_3$, $\sigma_4$, and $\sigma_5$, therefore, must be zero in the film so that we have

$$\phi \sigma_3' + (1-\phi)\sigma_3'' = 0, \quad \phi \sigma_4' + (1-\phi)\sigma_4'' = 0,$$
$$\phi \sigma_5' + (1-\phi)\sigma_5'' = 0. \qquad (10)$$

The local electric fields $\mathbf{E}'$ and $\mathbf{E}''$ in a polydomain ferroelectric film are not necessarily equal to the external field $\mathbf{E}_0$ defined by the potential difference between the electrodes. This effect may be caused by the presence of polarization charges at the film surfaces and even on domain boundaries. (We do not consider here films with other sources of internal fields, like charged vacancies, point defects, and depletion

layers.[46]) The surface depolarization field, however, may be ignored in relatively thick ($H$ > 100 nm) films of perovskite ferroelectrics, which are discussed in this paper. Indeed, the calculations taking into account the actual finite conductivity of these ferroelectrics show that the film depolarizing field is negligible.[47] This theoretical prediction is supported by the observations, which show that the domain structure of $PbZr_{0.2}Ti_{0.8}O_3$ epitaxial thin films is insensitive to the presence of electrodes.[18] Therefore, the mean electric field <**E**> in the film may be set equal to the applied field $\mathbf{E}_0$. This condition yields

$$\phi \mathbf{E}' + (1-\phi)\mathbf{E}'' = \mathbf{E_0}, \tag{11}$$

where it is implied that $\mathbf{E}_0$ is uniform, as in a conventional plate-capacitor setup.

In addition to the nine macroscopic conditions, which are described by Eqs. (3), (10), and (11), we can introduce "microscopic" boundary conditions on the domain walls. Indeed, the fields existing in adjacent domains are interrelated in the following way. First, the lattice strains in the polydomain layer must obey the classical compatibility condition.[48] In the rotated coordinate system ($x_1', x_2', x_3'$) with the $x_3'$ axis orthogonal to the walls, this requirement gives

$$S'_{1'} = S''_{1'}\ ,\quad S'_{2'} = S''_{2'}\ ,\quad S'_{6'} = S''_{6'}\ . \tag{12}$$

Second, from the equations of mechanical equilibrium,[49] written for a continuous medium in the absence of body forces, it follows that

$$\sigma'_{3'} = \sigma''_{3'},\quad \sigma'_{4'} = \sigma''_{4'},\quad \sigma'_{5'} = \sigma''_{5'}. \tag{13}$$

Third, the continuity of the tangential components of the internal electric field and the normal component of the electric displacement yields

$$E'_{1'} = E''_{1'}\ ,\quad E'_{2'} = E''_{2'}\ ,\quad \varepsilon_0 E'_{3'} + P'_{3'} = \varepsilon_0 E''_{3'} + P''_{3'}\ . \tag{14}$$

The eighteen relationships given by Eqs. (3) and (10)-(14) make it possible to express internal stresses $\sigma_i', \sigma_i''$ and electric fields $E_i', E_i''$ inside domains of two types in terms of polarization components $P_i', P_i''$ and the relative domain population $\phi$. After the substitution of these expressions into Eq. (2), the average energy density $<\tilde{F}>$ becomes a function of seven variables: $P_i', P_i''$ ($i$ = 1,2,3), and $\phi$. Performing numerically the minimization of $<\tilde{F}>(P_i', P_i''$, $\phi$), we can find the equilibrium polarizations in both domains and the equilibrium domain population $\phi$*. As follows from Eqs. (2), (3), and (11), these parameters depend of the misfit strain $S_m$, temperature $T$, and the applied electric field $\mathbf{E}_0$. The minimum energy $<\tilde{F}>$*($S_m$, $T$, $\mathbf{E}_0$) of the polydomain state can be also determined as a function of $S_m$, $T$, and $\mathbf{E}_0$.

On the basis of these calculations, the complete thermodynamic description of dense polydomain states in single crystalline films may be developed. This is a complicated task because several domain configurations, which differ by the spatial orientation of domain walls, are possible in epitaxial films. The theoretical analysis must include the comparison of the energies $<\tilde{F}>$* of different polydomain states with each other and with the energies of single-domain states, which can be also calculated with the aid of our theory. On this basis, the stability ranges of various thermodynamic states in the misfit strain-temperature plane may be determined for short-circuited films ($\mathbf{E}_0 = 0$). Using the constructed ($S_m$, $T$)-phase diagrams, the small-signal dielectric and piezoelectric constants of ferroelectric films may be computed as functions of the misfit strain and temperature. Moreover, the electric-field dependence of the film average polarization and its material constants can be studied, as well as the field-induced structural transformations in an epitaxial layer.

## III. PHASE DIAGRAMS OF $PbTiO_3$ AND $BaTiO_3$ EPITAXIAL THIN FILMS

For $PbTiO_3$ (PT) and $BaTiO_3$ (BT) films, quantitative results may be obtained with the aid of the procedure described in Sec. II, because the material parameters involved in the thermodynamic calculations are known for PT and BT to a good degree of precision. Using the values[50] of these parameters taken from Refs. 40-42, we have performed necessary numerical calculations and developed the misfit strain-temperature phase diagrams of short-circuited ($\mathbf{E}_0 = 0$) PT and BT films, which will be described in this section.

The following three variants of the orientation of domain walls were assumed to be possible in epitaxial films of perovskite ferroelectrics grown on cubic substrates.

(i) Domain walls are parallel to the {101} crystallographic planes of the prototypic cubic phase so that they are inclined at about 45° to the film/substrate interface $\Sigma$. This variant of domain geometry corresponds to the so-called *c*/*a*/*c*/*a* structure widely observed in epitaxial films of perovskite ferroelectrics (Fig. 1a).[2,6,10,11] It is composed of alternating tetragonal *c* domains and pseudo-tetragonal *a* domains, where the spontaneous polarization $\mathbf{P}_s$ is orthogonal to the interface $\Sigma$ in the *c* domains and parallel to $\Sigma$ in the *a* domains, being directed along the [100] axis of the prototypic phase in the latter.

(ii) Domain walls are orthogonal to the film/substrate interface and oriented along the {110} planes of the prototypic cubic phase. This domain-wall orientation is characteristic of the so-called $a_1/a_2/a_1/a_2$ structure, where the spontaneous polarization $\mathbf{P}_s$ develops in the film plane along the [100] and [010] axes within the $a_1$ and $a_2$ domains, respectively (Fig. 1b). Domain walls with this orientation were observed experimentally in $PbTiO_3$ epitaxial films.[6,8]

(iii) The walls are taken to be parallel to the {100} or {010} planes of the prototypic lattice, as expected for the domain patterning in the orthorhombic *aa* phase, which was predicted to form in single-domain PT and BT films at positive misfit strains.[38] The corresponding polydomain state may consist of orthorhombic $aa_1$ and $aa_2$ domains with $\mathbf{P}_s$ directed along the [110] and $[1\bar{1}0]$ axes of the cubic lattice, respectively (see Fig. 1c). In this $aa_1/aa_2/aa_1/aa_2$ structure, domain walls are also perpendicular to the substrate surface, but their orientation in the film plane differs from the preceding variant by 45°.

For each of the above three orientations of domain walls, the energetically most favorable polarization configurations at various temperatures $T$ and misfit strains $S_m$ were determined, and the minimum energies $<\tilde{F}>^*(S_m, T, \mathbf{E}_0 = 0)$ of corresponding polydomain states were evaluated and compared. The comparison was also made with the energy of the paraelectric phase ($P_1 = P_2 = P_3 = 0$) and with the minimum energies of homogeneous ferroelectric states possible in PT and BT epitaxial films, i.e. the *c* phase ($P_1 = P_2 = 0$, $P_3 \neq 0$), the *ca* phase (($P_1 \neq 0$, $P_2 = 0$, $P_3 \neq 0$), the *aa* phase ($P_1 = P_2 \neq 0$, $P_3 = 0$), and the *r* phase ($P_1 = P_2 \neq 0$, $P_3 \neq 0$).[38] Selecting then the energetically most favorable thermodynamic state for each point of the misfit strain-temperature plane, we obtained the equilibrium phase diagrams of PT and BT epitaxial films shown in Fig. 2.

Let us discuss first the diagram of $PbTiO_3$ films. At negative misfit strains $S_m$, except for a narrow strain range in the vicinity of $S_m = 0$, the paraelectric to ferroelectric transformation results in the appearance of the tetragonal *c* phase with the spontaneous polarization $\mathbf{P}_s$ orthogonal to the substrate surface. This result agrees with the earlier theoretical prediction.[38] Compressive in-plane stresses $\sigma_1$ and $\sigma_2$ in the ferroelectric phase prevent the film from twinning at negative misfit strains $S_m$ and relatively high temperatures $T$. During the further cooling of the epitaxial system, however, the introduction of elastic *a* domains into the *c* phase becomes energetically favorable. This leads to the formation of the pseudo-tetragonal *c*/*a*/*c*/*a* polydomain state with the standard "head-to-tail" polarization configuration and 90° walls (Fig. 3a).

As follows from our calculations, the spontaneous polarization $\mathbf{P}_s$ has the same magnitude in the *c* and *a* domains and varies with the misfit strain and temperature according to the relation

$$P_s^2 = -\frac{\alpha_{33}^{**}}{3\alpha_{111}} + \left( \frac{\alpha_{33}^{**2}}{9a_{111}^2} - \frac{\alpha_3^{**}}{3\alpha_{111}} \right)^{1/2}, \qquad (15)$$

where $\alpha_3^{**} = \alpha_1 - (Q_{12}/s_{11})S_m$ and $\alpha_{33}^{**} = \alpha_{11} + Q_{12}^2/2s_{11}$ are the renormalized coefficients of the second-order polarization term $P_3^2$ and the fourth-order term $P_3^4$ in the free-energy expansion (2). Internal electric fields and the stresses $\sigma_3$, $\sigma_4$, $\sigma_5$, and $\sigma_6$ are absent in the *c*/*a*/*c*/*a* structure. The equilibrium fraction of *c* domains equals

$$\phi_c^* = 1 - \frac{(s_{11} - s_{12})(S_m - Q_{12}P_s^2)}{s_{11}(Q_{11} - Q_{12})P_s^2}, \qquad (16)$$

and, at $\phi_c = \phi_c^*$, the stresses $\sigma_1^a$ and $\sigma_1^c$ also vanish, whereas the stress $\sigma_2$ acquires the same value of $\sigma_2^a = \sigma_2^c = (S_m - Q_{12}P_s^2)/s_{11}$ in both domains. The free-energy density $<\tilde{F}>*(S_m, T, \mathbf{E}_0 = 0)$ in the equilibrium *c*/*a*/*c*/*a* polydomain state may be written as

$$<\tilde{F}>* = \frac{S_m^2}{2s_{11}} + \alpha_3^{**}P_s^2 + \alpha_{33}^{**}P_s^4 + \alpha_{111}P_s^6 \qquad (17)$$

with $P_s^2$ given by Eq. (15).

Consider now the stability range $R_d$ of the *c*/*a*/*c*/*a* domain pattern in the ($S_m$, $T$)-phase diagram. The boundary of $R_d$, which is shown by a thin line in Fig. 2a, is defined by the inequality $\phi_c^*(S_m) < 1$ imposed on the equilibrium volume fraction of *c* domains. From Eq. (16) it follows that $\phi_c^*$ becomes equal to unity at $S_m^0(T) = Q_{12}P_0^2(T)$, where $P_0$ is the spontaneous polarization of a free crystal. The line $S_m^0(T) = Q_{12}[-\alpha_{11} + \sqrt{\alpha_{11}^2 - 3\alpha_1(T)\alpha_{111}}\,]/3\alpha_{111}$ constitutes the left-hand boundary of $R_d$, which is limited by the upper point of $R_d$ located at temperature $T_{\max} = \theta + 2\varepsilon_0 C\alpha_{11}^2/3\alpha_{111} \approx 497^0\mathrm{C}$ and misfit strain $S_m^0(T_{\max}) = Q_{12}(-\alpha_{11}/3\alpha_{111}) \approx 2.4\times10^{-3}$ (see Fig. 2a). The high-temperature section of right-hand boundary, which adjoins this upper point, is defined by the curve $S_m^0(T) = Q_{12}[-\alpha_{11} - \sqrt{\alpha_{11}^2 - 3\alpha_1(T)\alpha_{111}}\,]/3\alpha_{111}$. This part of $S_m^0(T)$ relates to the second possible solution for $P_0^2$ that exists at $T > \theta$ in crystals with $\alpha_{11} < 0$ and $\alpha_{111} > 0$.

The next section of the right-hand boundary of $R_d$ is formed by a short segment of the straight line, above which the solution (15) for the polarization in the *c*/*a*/*c*/*a* state loses its physical meaning (here $\alpha_{33}^{**2} < 3\alpha_3^{**}\alpha_{111}$). At the part of this segment situated between two triple points (see Fig. 2a), the direct transformation of the paraelectric phase into the polydomain *c*/*a*/*c*/*a* state takes place at the temperature

$$T_{c/a}(S_m) = \theta + \frac{2}{3}\varepsilon_0 C\frac{\alpha_{33}^{**2}}{\alpha_{111}} + 2\varepsilon_0 C\frac{Q_{12}}{s_{11}}S_m. \qquad (18)$$

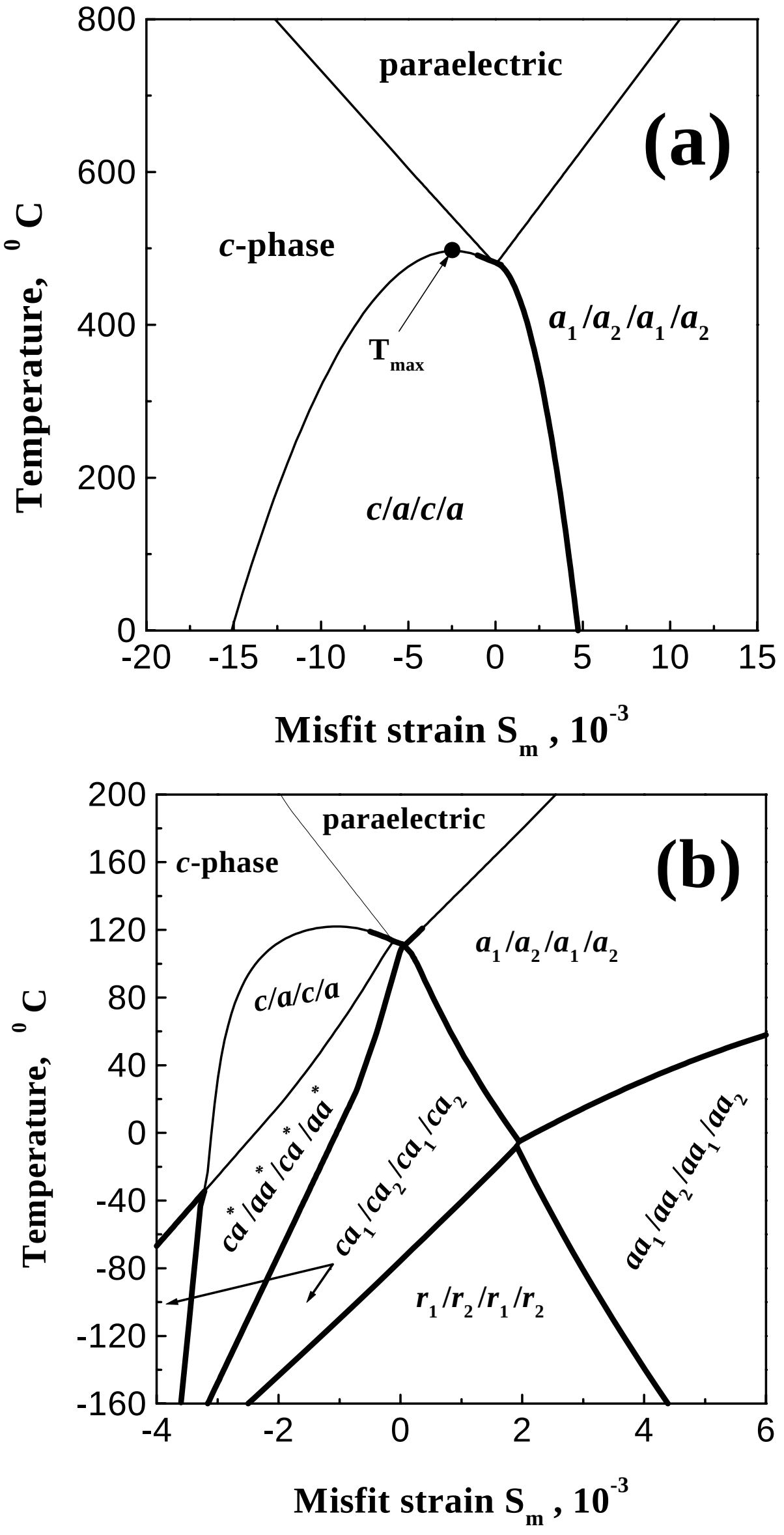

**FIG. 2.** Phase diagrams of $PbTiO_3$ (a) and $BaTiO_3$ (b) epitaxial films grown on cubic substrates. The continuous and discontinuous transformations are shown by thin and thick lines, respectively.

The ferroelectric phase transition in this “misfit-strain window” near $S_m = 0$ is of the *first order*, in contrast to the rest of the upper transition line, where it is of the *second order*. This is due to the fact that in PT (and also in BT) the renormalized coefficient $\alpha_{33}^{**}$ is *negative*. The above result corrects the earlier prediction[38] that in epitaxial PT and BT films the ferroelectric phase transition is expected to be always of the second order.

The remaining part of the right-hand boundary of $R_d$, also shown by a thick line in Fig. 2a, separates the stability ranges of the *c*/*a*/*c*/*a* and $a_1/a_2/a_1/a_2$ domain patterns in the ($S_m$, $T$)-plane. At larger positive misfit strains, the $a_1/a_2/a_1/a_2$ polarization configuration becomes the most energetically favorable thermodynamic state in PT films. The instability of the $c/a/c/a$ pattern with respect to the appearance of the polarization component $P_2$ parallel to the domain walls, which was described in Ref. 36, does not manifest itself in the diagram of *equilibrium states*. This $P_2$-instability occurs in PT films at positive misfit strains $S_m^*(T)$ outside $R_d$. However, the $P_2$-instability may be revealed by preparing a film with the equilibrium $c/a/c/a$ structure first and then bending the substrate to increase $S_m$ above $S_m^*(T)$ at a low temperature, where the formation of the $a_1/a_2/a_1/a_2$ state is suppressed.[36]

In the $a_1/a_2/a_1/a_2$ domain structure, the geometry of polarization patterning corresponds to the "head-to-tail" polarization configuration observed in bulk perovskite crystals.[40] The spontaneous polarization $\mathbf{P}_s$ has the same magnitude in the $a_1$ and $a_2$ domains and lies along the edges of the prototypic cubic cell, which are parallel to the film surfaces (see Fig. 3c). The equilibrium volume fractions of the $a_1$ and $a_2$ domains were found to be equal to each other at all investigated misfit strains and temperatures ($\phi^*$ = 0.5), which agrees with the result obtained earlier in the linear elastic approximation.[23,26,29] Internal electric fields and the stress components $\sigma_3$, $\sigma_4$, $\sigma_5$, and $\sigma_6$ are absent in the $a_1/a_2/a_1/a_2$ structure. The stresses $\sigma_1$ and $\sigma_2$ are homogeneous inside the film and, at the equilibrium domain population of $\phi^*$ = 0.5, acquire the same value of $\sigma_1 = \sigma_2 = [S_m - 0.5(Q_{11}+Q_{12})P_s^2]/(s_{11}+s_{12})$. In the rotated coordinate system, where the $x_1'$ axis is oriented along domain walls, the shear stress $\sigma_{6'}$ vanishes at $\phi^*$ = 0.5. This demonstrates a decrease in the film elastic energy, which is caused by the domain formation.

The magnitude of spontaneous polarization in the $a_1$ and $a_2$ domains depends on the misfit strain and temperature and can be found from the relation

$$P_s^2 = -\frac{\alpha_{11}^{**}}{3\alpha_{111}} + \left(\frac{\alpha_{11}^{**2}}{9\alpha_{111}^2} - \frac{\alpha_1^*}{3\alpha_{111}}\right)^{1/2}, \qquad (19)$$

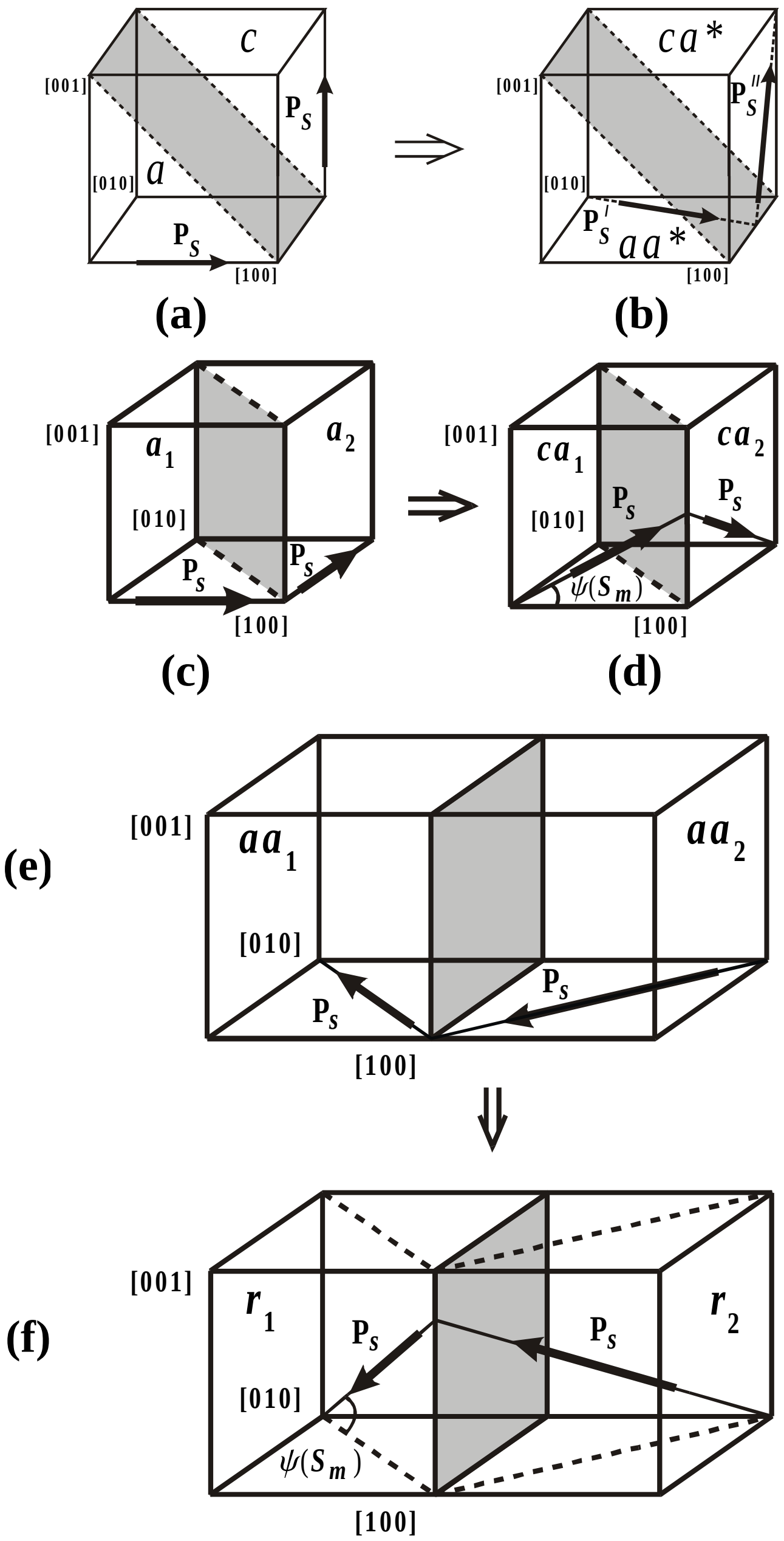

**FIG. 3.** Polarization patterning in various polydomain states forming in ferroelectric films: $c/a/c/a$ (a), $ca^*/aa^*/ca^*/aa^*$ (b), $a_1/a_2/a_1/a_2$ (c), $ca_1/ca_2/ca_1/ca_2$ (d), $aa_1/aa_2/aa_1/aa_2$ (e), $r_1/r_2/r_1/r_2$ (f). Polarization orientations are shown relative to the prototypic cubic cell.

where $\alpha_1^* = \alpha_1 - (Q_{11}+Q_{12})S_m/(s_{11}+s_{12})$ is the coefficient introduced in Ref. 38, and $\alpha_{11}^{**} = \alpha_{11} + (Q_{11}+Q_{12})^2/4(s_{11}+s_{12})$ is the renormalized coefficient of the fourth-order polarization term ($P_1^4 + P_2^4$) in the free-energy expansion (2), which differs from the similar coefficient $\alpha_{11}^*$ appearing in the theory of single-domain films.[38] The mean energy density $<\tilde{F}>^*$($S_m$, $T$, $\mathbf{E}_0$ = 0) stored in the film with an equilibrium $a_1/a_2/a_1/a_2$ structure is given by

$$< \tilde{F} >* = \frac{S_m^2}{s_{11}+s_{12}} + \alpha_1^* P_s^2 + \alpha_{11}^{**} P_s^4 + \alpha_{111} P_s^6 \qquad (20)$$

with $P_s^2$ defined by Eq. (19).

From comparison of Fig. 2a with the phase diagram of single-domain PT films[38] it follows that the $a_1/a_2/a_1/a_2$ polydomain state replaces the orthorhombic *aa* phase in the equilibrium diagram. Nevertheless, the paraelectric to ferroelectric phase transition in our approximation remains to be of the *second order* at positive misfit strains $S_m$ (except for a narrow range near $S_m = 0$, where the *c*/*a*/*c*/*a* state forms). This is due to the fact that the renormalized coefficient $\alpha_{11}^{**}$ is *positive* in PT films, as well as $\alpha_{11}^*$.

Proceed now to the equilibrium phase diagram of BT films. From inspection of Fig. 2 it can be seen that it has much more complicated structure than the diagram of PT films. The analysis shows that this is due to the existence of three different polarization instabilities in epitaxial BT films. The first one is the $P_2$-instability of the *c*/*a*/*c*/*a* polydomain state described in Ref. 36.[51] In contrast with PT films, where this instability appears only in metastable *c*/*a*/*c*/*a* structures, in BT films it manifests itself in the diagram of equilibrium thermodynamic states leading to the formation of the heterophase *ca**/*aa**/*ca**/*aa** state (see Fig. 3d) composed of distorted *ca* and *aa* phases.[36]

The second instability refers to the pseudo-tetragonal $a_1/a_2/a_1/a_2$ state. Under certain $S_m$-$T$ conditions, this state becomes unstable with respect to the *in-plane rotation* of $\mathbf{P}_s$ away from the edges of the prototypic cubic cell, which leads to its transformation into the orthorhombic *aa* phase. This instability appears during the film cooling, when the misfit strain $S_m$ in the epitaxial system is larger than about $2\times10^{-3}$. If the misfit strain increases at a given temperature due to the substrate bending, for example, the transformation also should take place at some critical value of $S_m$ (about $5.9\times10^{-3}$ at 25°C). This behavior may be explained in a natural way because in stress-free BT crystals at temperatures below 10°C the orthorhombic phase becomes energetically more favorable than the tetragonal one.[40]

The homogeneous *aa* phase, however, does not appear in the diagram of equilibrium states. The calculations show that it always tend to convert into the polydomain $aa_1/aa_2/aa_1/aa_2$ state, which is energetically more favorable. As a result, well below the transition line $T_c(S_m > 0) = T_{aa}(S_m)$, the $aa_1/aa_2/aa_1/aa_2$ domain pattern replaces the $a_1/a_2/a_1/a_2$ one in the equilibrium diagram of BT films (see Fig. 2b). In the $aa_1/aa_2/aa_1/aa_2$ state, the polarization patterning occurs along two in-plane *face diagonals* of the prototypic cubic cell (Fig. 3e) so that 90° domain walls form here. The equilibrium volume fractions of the $aa_1$ and $aa_2$ domains are equal to each other ($\phi^* = 0.5$), and the spontaneous polarization $\mathbf{P}_s$ has the same magnitude in these domains. Internal electric fields and the stresses $\sigma_3$, $\sigma_4$, $\sigma_5$, and $\sigma_6$ are absent in the equilibrium $aa_1/aa_2/aa_1/aa_2$ structure. The stresses $\sigma_1$ and $\sigma_2$ are homogeneous inside the film and, in similarity with the $a_1/a_2/a_1/a_2$ state, have the same value of $\sigma_1 = \sigma_2 = [S_m - 0.5(Q_{11}+Q_{12})P_s^2]/(s_{11}+s_{12})$. The spontaneous polarization can be calculated from the formula

$$P_s^2 = -\frac{2a_{11}^* + a_{12}^{**}}{3(a_{111}+a_{112})} + \left[\frac{(2a_{11}^* + a_{12}^{**})^2}{9(a_{111}+a_{112})^2} - \frac{4a_1^*}{3(a_{111}+a_{112})}\right]^{1/2}, \qquad (21)$$

where $\alpha_{12}^{**} = \alpha_{12}^* - Q_{44}^2/(2s_{44})$, and $\alpha_{11}^*$ and $\alpha_{12}^*$ are the renormalized coefficients of the free-energy expansion, which were introduced in Ref. 38.[52] The mean energy density $< \tilde{F} >*(S_m, T, \mathbf{E}_0 = 0)$ in the equilibrium $aa_1/aa_2/aa_1/aa_2$ polydomain state can be found from the relation

$$< \tilde{F} >* = \frac{S_m^2}{s_{11}+s_{12}} + \alpha_1^* P_s^2 + \frac{1}{4}(2a_{11}^* + a_{12}^{**})P_s^4 + \frac{1}{4}(a_{111}+a_{112})P_s^6, \qquad (22)$$

where $P_s^2$ is given by Eq. (21).

The third instability, which exists in epitaxial BT films, is the instability of the $a_1/a_2/a_1/a_2$ and $aa_1/aa_2/aa_1/aa_2$ states with respect to the *appearance* of the polarization component $P_3$ orthogonal to the film surfaces. This $P_3$-instability occurs below the Curie-Weiss temperature $\theta$ = 108 $^0$C of bulk BT crystals, when the magnitude of the positive misfit strain in the film/substrate system *decreases* down to some critical value. It also appears during the film cooling under certain misfit-strain conditions.

Owing to the $P_3$-instability, the $a_1/a_2/a_1/a_2$ and $aa_1/aa_2/aa_1/aa_2$ configurations transform into *new* polydomain states, which may be termed $ca_1/ca_2/ca_1/ca_2$ and $r_1/r_2/r_1/r_2$ structures, respectively. In both states, the out-of-plane polarizations in neighboring domains have opposite directions but the same magnitude ($P_3' = -P_3''$), whereas the in-plane polarizations retain the orientations characteristic of the $a_1/a_2/a_1/a_2$ and $aa_1/aa_2/aa_1/aa_2$ configurations (see Fig. 3d,f). Since in the $ca_1/ca_2/ca_1/ca_2$ pattern the spontaneous polarizations are parallel to the out-of-plane *faces* of the prototypic cubic cell, it may be regarded as a polydomain analogue of the monoclinic *ca* phase forming in single domain BT films.[38] In the $r_1/r_2/r_1/r_2$ state, the in-plane polarizations are oriented along the face *diagonals* of the prototypic cell so that the spatial orientation of $\mathbf{P}_s$ here is similar to that in the homogeneous monoclinic *r* phase.[38] The equilibrium volume fractions of the $r_1$ and $r_2$ domains are equal to each other ($\phi$ * = 0.5), as well as the populations of the $ca_1$ and $ca_2$ domains.

Remarkably, the spatial orientations of the spontaneous polarization $\mathbf{P}_s$ in the $ca_1/ca_2/ca_1/ca_2$ and $r_1/r_2/r_1/r_2$ states depend on the misfit strain $S_m$ and temperature *T*. Therefore, the vector $\mathbf{P}_s$ generally is not parallel to the face or cube diagonals of the prototypic cell (Fig. 3d,f) so that these two states have no direct analogue in bulk BT crystals. The domain walls in these patterns are not 90° walls anymore. Since the rotation of $\mathbf{P}_s$ at the $ca_1/ca_2$ wall is larger than 90° but generally not equal to 120°, this wall differs from 120° domain boundaries in the orthorhombic phase of a bulk BT crystal. The $r_1/r_2$ walls are not equivalent to the domain walls in the stress-free rhombohedral BT crystal as well. Nevertheless, the presence of the $r_1/r_2/r_1/r_2$ pattern in the ($S_m$, *T*)-diagram has probably the same origin as the formation of the rhombohedral phase in bulk BT crystals at temperatures below –71°C.[40-41]

Thus, in BT films the stability range of polydomain states with walls orthogonal to the substrate surface splits into four parts. These parts are separated by the first-order transition lines and correspond to the $a_1/a_2/a_1/a_2$, $aa_1/aa_2/aa_1/aa_2$, $ca_1/ca_2/ca_1/ca_2$, and $r_1/r_2/r_1/r_2$ configurations.

Finally, it should be noted that there are some similarities between the phase diagrams of BT and PT films. For example, the transition line between the homogeneous *c* phase and the polydomain *c*/*a*/*c*/*a* state is defined in BT by the relationships discussed above for PT films.

## IV. DIELECTRIC PROPERTIES OF POLYDOMAIN FERROELECTRIC FILMS

In general, the dielectric response of a polydomain or heterophase film is a sum of *intrinsic* and *extrinsic* contributions. Changes of the polarizations $\mathbf{P}'$, $\mathbf{P}''$ inside dissimilar domains (phase layers) lead to an average intrinsic (volume) response. If the measuring field $\mathbf{E}_0$ also induces reversible displacements of domain walls (phase boundaries) from their initial equilibrium positions, an additional extrinsic contribution may appear, being caused by rotations of the polarization vector in the part of the film volume swept by moving walls.[27,28] Fortunately, our thermodynamic theory makes it possible to calculate the *total permittivity* of a polydomain or heterophase film, which results from both intrinsic and extrinsic contributions. To that end, equilibrium polarizations $\mathbf{P}'$, $\mathbf{P}''$ in the domains (phase layers) of two types and their equilibrium volume fractions should be determined numerically as a function of the external electric field $\mathbf{E}_0$. Then the permittivity of a ferroelectric film can be found from the field dependence of the average polarization $<\mathbf{P}> = \phi^* \mathbf{P}' + (1-\phi^*)\mathbf{P}''$ in an epitaxial layer. Choosing a reasonable magnitude $E_0$ of the weak external field, we calculated the small-signal dielectric constants $\varepsilon_{ij}$ ($i, j$ = 1,2,3) of PT and BT films from the relation

$$\varepsilon_{ij} = \frac{\langle P_i \rangle (E_{0j} = E_0) - \langle P_i \rangle (E_{0j} = 0)}{E_0}. \qquad (23)$$

Since the equilibrium domain population $\phi$* was allowed to vary under field $\mathbf{E}_0$, the constants $\varepsilon_{ij}$ involved a nonzero extrinsic (domain-wall) contribution when the projection of the spontaneous polarization on $\mathbf{E}_0$ was different in dissimilar domains. The calculated constants also correctly take into account the influence of the mechanical film/substrate interaction on the intrinsic (volume) responses of the domains of two types.

Figure 4 shows variations of the diagonal components of the dielectric tensor $\varepsilon_{ij}$ with the misfit strain $S_m$ in PT films at room temperature. At the transition point $S_m^0(T = 25°C) =$ -14.7×10$^{-3}$, where the transformation of the *c* phase into the *c*/*a*/*c*/*a* pattern takes place, finite jumps of $\varepsilon_{11}$ and $\varepsilon_{33}$ occur, while the dependence $\varepsilon_{22}(S_m)$ only changes its slope. The analysis shows that the step-like increase of $\varepsilon_{11}$ and $\varepsilon_{33}$ at $S_m = S_m^0$ is entirely due to the appearance of domain-wall contributions to these dielectric constants in the polydomain *c*/*a*/*c*/*a* state. If domain walls are assumed to be pinned so that the domain population $\phi(\mathbf{E}_0)$ is kept fixed at a value of $\phi^*(\mathbf{E}_0 = 0)$ during the calculations, all constants $\varepsilon_{ii}$ vary in a continuous manner on crossing the transition line (see dashed lines in Fig. 4). The slope of $\varepsilon_{33}(S_m)$ changes slightly here, whereas $\varepsilon_{11}(S_m)$ has a cusp-like anomaly at $S_m = S_m^0$. Such dielectric behavior can be attributed to the fact that the formation of the polydomain *c*/*a*/*c*/*a* state in our approximation proceeds via the introduction of a small volume fraction of the *a* domains (with a finite spontaneous polarization $P_0$) into the *c* phase.

It should be noted that the theoretical domain-wall contribution $\Delta\varepsilon_{33}/\varepsilon_0$ to the film permittivity $\varepsilon_{33}$, which is measured in a conventional plate-capacitor setup, increases from 70 to 120 in the range of misfit strains between –14.7×10$^{-3}$ and 4.6×10$^{-3}$, where the *c*/*a*/*c*/*a* pattern represents the most energetically favorable configuration. These values may be compared with the results of approximate analytical calculations of $\Delta\varepsilon_{33}$, which can be performed in the framework of our theory as well. In the first approximation, the effect of small domain-wall displacements on the polarizations $\mathbf{P}'$, $\mathbf{P}''$ inside *c* and *a* domains may be neglected. Evaluating in this approximation the field-induced change of the domain population $\phi$

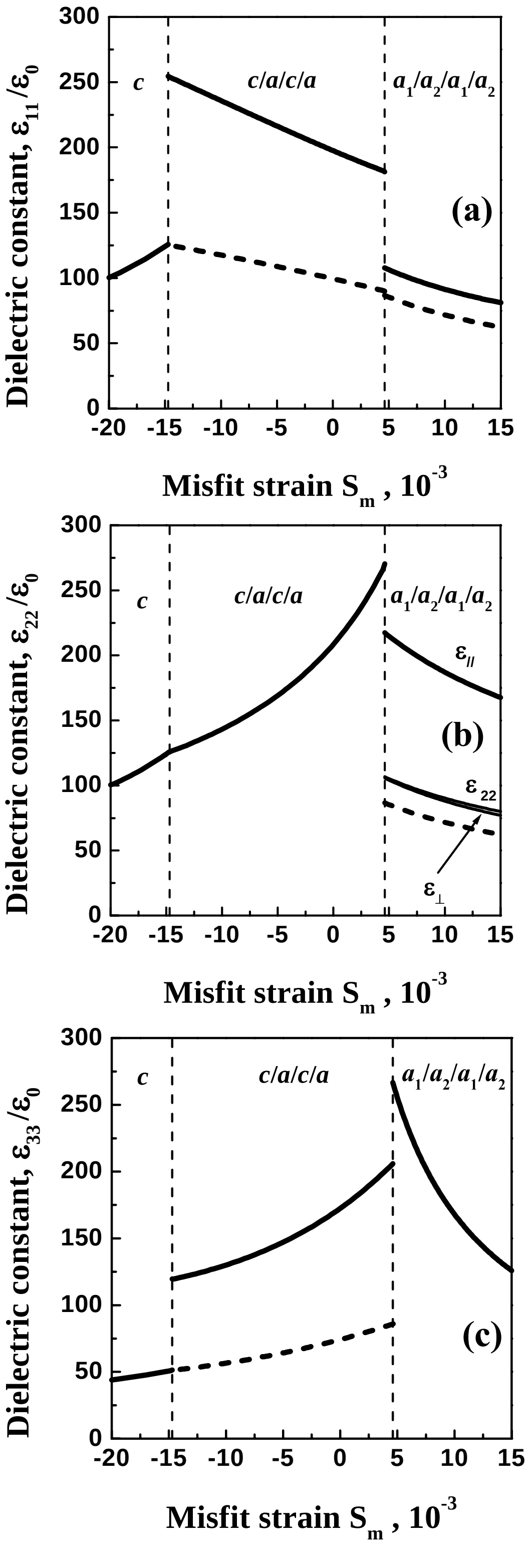

**FIG. 4.** Dependencies of the dielectric constants $\varepsilon_{11}$ (a), $\varepsilon_{22}$ (b), and $\varepsilon_{33}$ (c) of $PbTiO_3$ films on the misfit strain $S_m$ at $T$ = 25°C. The dashed lines show the permittivities of $PbTiO_3$ films with pinned domain walls. For films with the $a_1/a_2/a_1/a_2$ domain structure, the in-plane dielectric responses $\varepsilon_{\parallel}$ and $\varepsilon_{\perp}$ in the directions parallel and orthogonal to the 90$^0$ walls are also shown for comparison.

via the mathematical procedure developed in Sec. II, from Eq. (23) we obtain

$$\Delta\varepsilon_{33} \approx \frac{\left(s_{11}^2 - s_{12}^2\right)}{s_{11}\left(Q_{11} - Q_{12}\right)^2 P_s^2}. \quad (24)$$

Equation (24) is similar to an analytic expression, which was derived for $\Delta\varepsilon_{33}$ earlier in the approximation of the linear isotropic theory of elasticity.[27,28] The substitution of Eq. (15) for the spontaneous polarization $P_s$ into Eq. (24) gives $\Delta\varepsilon_{33}/\varepsilon_0 \approx 110\div130$ for PT films at room temperature, which is in reasonable agreement with our numerical results. Though the linear approximation overestimates $\Delta\varepsilon_{33}$, the main conclusion made in Refs. 27-28 is justified by the nonlinear theory. Indeed, according to our calculations, the domain-wall contribution $\Delta\varepsilon_{33}$ represents more than a half of the total response of a polydomain PT film so that translational displacements of 90° walls really contribute considerably to the film permittivity. (As demonstrated by Fig. 5c, a similar situation takes place in BT films with the equilibrium *c*/*a*/*c*/*a* domain structure. Here the domain-wall contribution $\Delta\varepsilon_{33}/\varepsilon_0$ varies between 375 and 387, which may be compared with $\Delta\varepsilon_{33}/\varepsilon_0 \approx 510\div530$ given by the linear theory.)

When the misfit strain in the heterostructure increases above a value of $S_m = 4.6\times10^{-3}$, the $a_1/a_2/a_1/a_2$ polydomain state replaces the *c*/*a*/*c*/*a* one in the phase diagram of PT films. At this threshold strain, as may be expected, the film dielectric constants $\varepsilon_{ii}$ exhibit step-like changes (see Fig. 4). Remarkably, in the vicinity of the threshold strain the permittivity $\varepsilon_{33}$ of an epitaxial PT film becomes considerably higher than the largest dielectric constant of a single-domain bulk PT crystal (about 125 at room temperature according to the thermodynamic calculations).

For films with the $a_1/a_2/a_1/a_2$ domain pattern, the dielectric response $\varepsilon_{33}$ in the film thickness direction can be calculated analytically. In this case the domain-wall contribution is absent and the inverse dielectric susceptibility $\chi_{33}$ equals

$$\chi_{33} = 2\alpha_3^* + 2\alpha_{13}^{**}P_s^2 + 2\alpha_{112}P_s^4, \quad (25)$$

where $\alpha_3^* = \alpha_1 - S_m 2Q_{12}/(s_{11}+s_{12})$ is the coefficient introduced in Ref. 38, and $\alpha_{13}^{**} = \alpha_{12} + (Q_{11}+Q_{12})Q_{12}/(s_{11}+s_{12}) + Q_{44}^2/(4s_{44})$ is the renormalized coefficient of the fourth-order polarization term $(P_1^2 + P_2^2)P_3^2$ in the free-energy expansion (2), which differs from the similar coefficient $\alpha_{13}^*$ appearing in the theory of single-domain ferroelectric films, i.e. $\alpha_{13}^{**} = \alpha_{13}^* + Q_{44}^2/(4s_{44})$. The substitution of Eq. (19) into Eq. (25) makes it possible to calculate $\varepsilon_{33} = \varepsilon_0 + \chi_{33}^{-1}$ as a function on the misfit strain and temperature.

The in-plane dielectric responses $\varepsilon_{11}$ and $\varepsilon_{22}$ of a film with the $a_1/a_2/a_1/a_2$ structure, measured along the [100] and [010] crystallographic axes, are equal to each other and contain a nonzero domain-wall contribution (see Figs. 4a and 4b). This contribution appears because the electric field is directed at 45° to the 90° walls (see Fig. 1b). For comparison, we also calculated the film in-plane permittivities $\varepsilon_\perp$ and $\varepsilon_\parallel$ in the directions orthogonal and parallel to the $a_1/a_2$ domain walls. When the measuring field $\mathbf{E}_0$ is orthogonal to the walls, there is no driving force acting on them and the domain-wall contribution to the corresponding permittivity $\varepsilon_\perp$ equals zero. In contrast, the field $\mathbf{E}_0$ oriented along $a_1/a_2$ walls in the film plane induces the wall displacements from equilibrium positions so that the dielectric response $\varepsilon_\parallel$ should contain a nonzero domain-wall contribution $\Delta\varepsilon_\parallel$. From inspection of Fig. 4b it can be seen that $\varepsilon_\parallel$ is about two times larger than $\varepsilon_\perp$.

Let us discuss now peculiarities of the dielectric properties of BT films. The misfit-strain dependences of the film permittivities $\varepsilon_{ii}$ at room temperature are shown in Fig. 5. As may be expected, the polarization instabilities of polydomain states, which were described in Sec. III, manifest itself in strong dielectric anomalies. The transition between the *c*/*a*/*c*/*a* and *ca**/*aa**/*ca**/*aa** states is accompanied by a singularity of $\varepsilon_{22}$, which is caused by the $P_2$-instability. A sharp peak of $\varepsilon_{33}$ is associated with the $P_3$-instability of the $a_1/a_2/a_1/a_2$ pattern, which leads to the formation of the $ca_1/ca_2/ca_1/ca_2$ configuration. On the other hand, the instability of the $a_1/a_2/a_1/a_2$ state with respect to the in-plane rotation of the spontaneous polarization does not manifest itself in Fig. 5. This is due to the fact that the critical misfit strain $S_m$ = 5.89×10$^{-3}$ for this instability is considerably larger than the strain $S_m$ = 3.65×10$^{-3}$, at which the $aa_1/aa_2/aa_1/aa_2$ pattern replaces the $a_1/a_2/a_1/a_2$ structure is the equilibrium phase diagram. At the transition between these two in-plane polarization states, the dielectric constants $\varepsilon_{ii}$ experience only step-like changes.

The dielectric response $\varepsilon_{33}$ of the film with the $aa_1/aa_2/aa_1/aa_2$ domain pattern can be found analytically, because it does not contain the domain-wall contribution. For the inverse dielectric susceptibility $\chi_{33}$, the calculation yields

$$\chi_{33} = 2\alpha_3^* + 2\alpha_{13}^{**}P_s^2 + (\alpha_{112} + \frac{1}{2}\alpha_{123})P_s^4, \qquad (26)$$

where $P_s^2$ given by Eq. ( 21) should be substituted. The permittivity $\varepsilon_{33} = \varepsilon_0 + \chi_{33}^{-1}$ increases with decreasing positive misfit strain, as demonstrated by Fig. 5c.

In conclusion of this section we analyze domain-wall contributions to the in-plane dielectric responses $\varepsilon_{11}$ and $\varepsilon_{22}$ of polydomain ferroelectric films, which could be measured in setups with interdigitated surface electrodes. For films with the *c*/*a*/*c*/*a* structure, the domain-wall contribution $\Delta\varepsilon_{11}$ calculated in the linear approximation is equal to the contribution $\Delta\varepsilon_{33}$ by reason of symmetry. Using Eq. (24), we obtain $\Delta\varepsilon_{11}/\varepsilon_0 \approx$ 110÷130 and 510÷530 for PT and BT films, respectively. These estimates are in good agreement with the results of our numerical calculations, which give $\Delta\varepsilon_{11}/\varepsilon_0$ between 130 and 90 for PT films and between 600 and 580 for BT ones (see Figs. 4a and 5a).

The domain-wall contributions to the dielectric constants of ferroelectric films with the $a_1/a_2/a_1/a_2$ and $aa_1/aa_2/aa_1/aa_2$ structures can be evaluated in the linear approximation by analogy with the derivation of Eq. (24). Neglecting polarization changes accompanying the field-induced displacements of 90$^0$ walls from their initial positions, we obtain

$$\Delta\varepsilon_{11} = \Delta\varepsilon_{22} \approx \frac{(s_{11} - s_{12})}{2(Q_{11} - Q_{12})^2 P_s^2} \qquad (27)$$

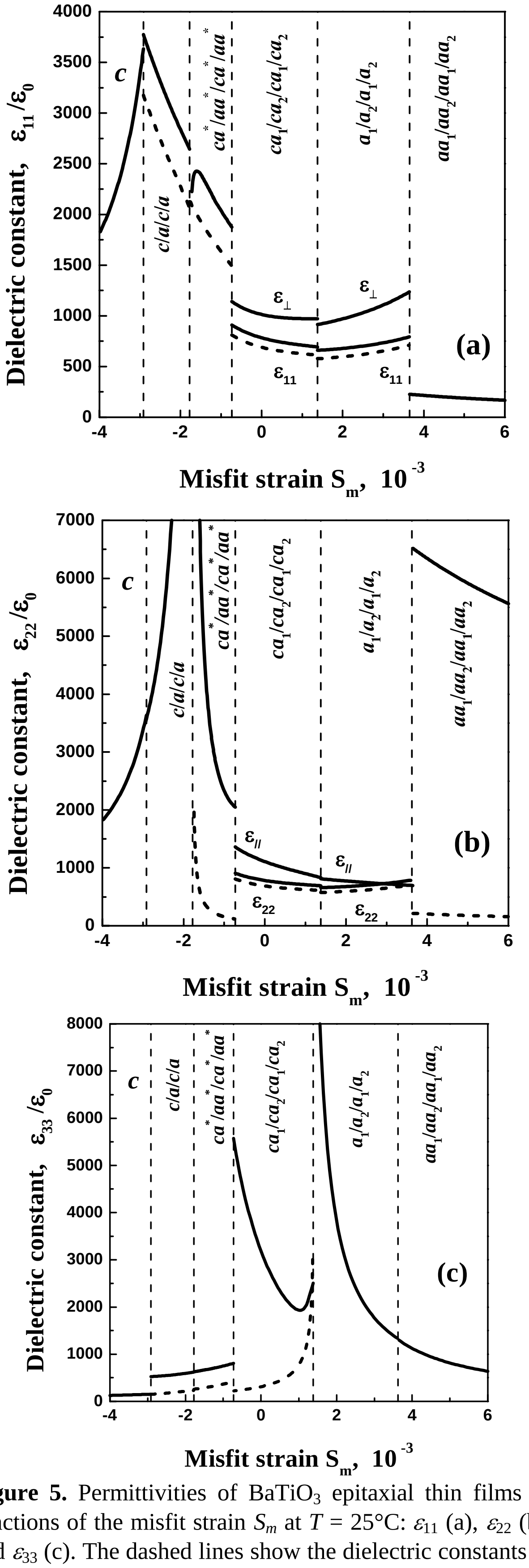

**Figure 5.** Permittivities of $BaTiO_3$ epitaxial thin films as functions of the misfit strain $S_m$ at $T$ = 25°C: $\varepsilon_{11}$ (a), $\varepsilon_{22}$ (b), and $\varepsilon_{33}$ (c). The dashed lines show the dielectric constants of $BaTiO_3$ films with pinned domain walls. For films with the $a_1/a_2/a_1/a_2$ and $ca_1/ca_2/ca_1/ca_2$ domain structures, the in-plane dielectric responses $\varepsilon_{\|}$ and $\varepsilon_{\perp}$ in the directions parallel and orthogonal to the domain walls are also shown for comparison.

for the $a_1/a_2/a_1/a_2$ state and

$$\Delta\varepsilon_{22} \approx \frac{2s_{44}}{Q_{44}^2 P_s^2} \quad , \tag{28}$$

for the $aa_1/aa_2/aa_1/aa_2$ one. The substitution of Eq. (21) for the spontaneous polarization $P_s$ into Eq. (28) shows that in BT films with the equilibrium $aa_1/aa_2/aa_1/aa_2$ structure $\Delta\varepsilon_{22}/\varepsilon_0 \approx 6500$ at the threshold misfit strain $S_m = 3.65\times10^{-3}$ in good agreement with the contribution $\Delta\varepsilon_{22}/\varepsilon_0 = 6300$ calculated numerically. In the case of the $a_1/a_2/a_1/a_2$ domain configuration, from Eqs. (19) and (27) we obtain $\Delta\varepsilon_{11}/\varepsilon_0 \approx 110$ for PT films at $S_m = 4.6\times10^{-3}$ and $\Delta\varepsilon_{11}/\varepsilon_0 \approx 400$ for BT films at $S_m = 1.38\times10^{-3}$, for example. According to our numerical calculations, the contribution $\Delta\varepsilon_{11}/\varepsilon_0$ equals 20 and 80 in these two situations, respectively. It can be seen that the linear approximation strongly overestimates the domain-wall contribution to the dielectric response in this particular case.

## V. PIEZOELECTRIC RESPONSE OF POLYDOMAIN FERROELECTRIC FILMS

In this section the piezoelectric properties of polydomain ferroelectric thin films epitaxially grown on thick cubic substrates will be described. We restrict our study to the converse piezoelectric effect, which consists in changes of the film dimensions and shape under the action of an external electric field $\mathbf{E}_0$.

The small-signal piezoelectric coefficients $d_{in}$ characterizing this effect can be found from the field dependence of the mean strains $\langle S_n \rangle = \phi^* S'_n + (1-\phi^*) S''_n$ in a polydomain film as

$$d_{in} = \frac{\langle S_n \rangle (E_{0i} = E_0) - \langle S_n \rangle (E_{0i} = 0)}{E_0}. \tag{29}$$

Since the epitaxial film is rigidly connected with a thick substrate, which is assumed to be piezoelectrically inactive, the strains $\langle S_1 \rangle$, $\langle S_2 \rangle$, and $\langle S_6 \rangle$ cannot change under the field so that $d_{i1} = d_{i2} = d_{i6} = 0$. Variations of the strains $\langle S_3 \rangle$, $\langle S_4 \rangle$, and $\langle S_5 \rangle$ manifest itself in a change of the film thickness and a tilt of the ferroelectric overlayer relative to the substrate normal. These strains can be expressed in terms of polarization components in the domains of two types and the relative domain population $\phi$ as

$$\langle S_3 \rangle = \frac{2s_{12}}{s_{11}+s_{12}} S_m + \frac{Q_{12}s_{11} - Q_{11}s_{12}}{s_{11}+s_{12}} \times \left[\phi\left(P_1'^2 + P_2'^2\right) + (1-\phi)\left(P_1''^2 + P_2''^2\right)\right] + \left(Q_{11} - \frac{2Q_{12}s_{12}}{s_{11}+s_{12}}\right)\left[\phi P_3'^2 + (1-\phi)P_3''^2\right], \tag{30}$$

$$\langle S_4 \rangle = Q_{44}\left[\phi P_2' P_3' + (1-\phi) P_2'' P_3''\right], \tag{31}$$

$$\langle S_5 \rangle = Q_{44}\left[\phi P_1' P_3' + (1-\phi) P_1'' P_3''\right]. \tag{32}$$

In a conventional plate-capacitor setup, where the film is sandwiched between two continuous electrodes, the piezoelectric measurements give the coefficients $d_{33}$, $d_{34}$, and $d_{35}$. Using Eq. (29), we calculated numerically these piezoelectric constants for PT and BT films with equilibrium domain structures. Figures 6 and 7 show the theoretical misfit-strain dependencies of $d_{33}$, $d_{34}$, and $d_{35}$ at room temperature.

From inspection of Fig. 6a it can be seen that the longitudinal piezoelectric coefficient $d_{33}$ of PT films increases with the increase of the misfit strain $S_m$ up to the threshold strain $S_m = 4.6\times10^{-3}$, at which the $c/a/c/a$ polydomain state is replaced by the $a_1/a_2/a_1/a_2$ one. A similar trend is characteristic of the misfit-strain dependence of $d_{33}$ in BT films, where the maximum response is displayed by a film with the $ca^*/aa^*/ca^*/aa^*$ structure (see Fig. 7a). It should be noted that these theoretical results correspond to the observed properties of prepolarized ferroelectric films, where all $180^0$ domain walls are removed prior to the piezoelectric measurements.

The step-like increase of the constant $d_{33}$ at the transition from the $c$ phase to the $c/a/c/a$ state is due to the appearance of a nonzero domain-wall contribution $\Delta d_{33}$ (see Figs. 6a and 7a). Within the stability range of the $c/a/c/a$ structure the value of $\Delta d_{33}$ varies from 17 to 50 pm/V in PT films and between 56 and 61 pm/V in BT films. These values may be compared with the domain-wall contribution $\Delta d_{33}$ calculated for dense domain structures with the aid of the linear theory.[31] Using the relation derived in Ref. 31 and taking into account the elastic anisotropy of the paraelectric phase, we obtain

$$\Delta d_{33} = \frac{(s_{11} + 2s_{12})(s_{11} - s_{12})}{s_{11}(Q_{11} - Q_{12})P_0}. \qquad (33)$$

For PT and BT films at room temperature, Eq. (33) gives $\Delta d_{33}$ about 45 and 95 pm/V, respectively. These values are in reasonable agreement with the aforementioned results of our calculations. Though the linear approximation overestimates $\Delta d_{33}$, the earlier prediction[31] that the displacements of $90^0$ domain walls may contribute considerably to the longitudinal piezoelectric response of polydomain ferroelectric thin films is confirmed by the nonlinear thermodynamic theory (see Figs. 6a and 7a).

The numerical calculations also show that ferroelectric films with the equilibrium $a_1/a_2/a_1/a_2$, $aa_1/aa_2/aa_1/aa_2$, and $ca_1/ca_2/ca_1/ca_2$ domain structures have a negligible piezoelectric coefficient $d_{33}$. When the field-induced displacements of domain walls are absent, as in the $a_1/a_2/a_1/a_2$ and $aa_1/aa_2/aa_1/aa_2$ states, from Eqs. (29)-(30) it follows that the piezoelectric response $d_{33}$ can be written as

$$d_{33} = 2\frac{Q_{12}s_{11} - Q_{11}s_{12}}{s_{11} + s_{12}} \times \left[\phi(\eta'_{13}P'_1 + \eta'_{23}P'_2) + (1-\phi)(\eta''_{13}P''_1 + \eta''_{23}P''_2)\right] + 2\left(Q_{11} - \frac{2Q_{12}s_{12}}{s_{11} + s_{12}}\right)\left[\phi\eta'_{33}P'_3 + (1-\phi)\eta''_{33}P''_3\right], \qquad (34)$$

where $\eta'_{ij}$ and $\eta''_{ij}$ are the intrinsic dielectric susceptibilities of the ferroelectric material inside domains of the first and second kind, respectively. Equation (34) demonstrates that the piezoelectric response $d_{33}$ of the $a_1/a_2/a_1/a_2$ and $aa_1/aa_2/aa_1/aa_2$ states must be really very small. Indeed, in these states at $\mathbf{E_0} = 0$ only the in-plane polarization differs from zero, which is multiplied by nondiagonal components of the dielectric tensor $\eta_{ij}$ in the expression for $d_{33}$. The fact that films with the $ca_1/ca_2/ca_1/ca_2$ structure have a negligible coefficient $d_{33}$ may be explained by the zero

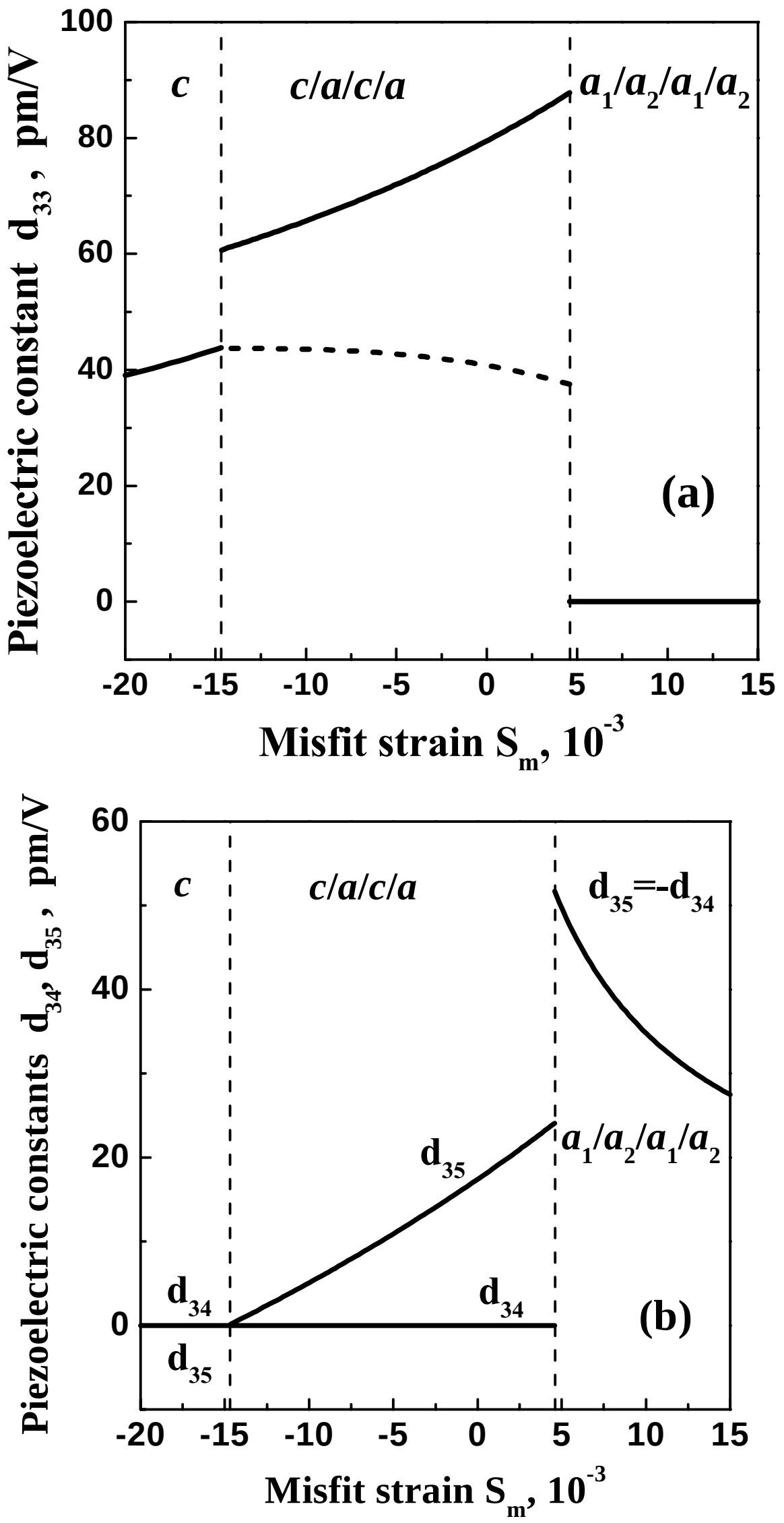

**FIG. 6.** Dependencies of the piezoelectric coefficients $d_{33}$ (a) and $d_{34}$, $d_{35}$ (b) of $PbTiO_3$ films on the misfit strain $S_m$ at $T$ = 25°C. The dashed line shows the piezoelectric response $d_{33}$ of $PbTiO_3$ films with pinned domain walls.

value of the average out-of-plane polarization $P_3$ in this state as well.

Consider now the misfit-strain dependencies of the piezoelectric coefficients $d_{34}$ and $d_{35}$ in epitaxial PT and BT films (Figs. 6b and 7b). By reason of symmetry $d_{34} = d_{35} = 0$ in the *c* phase and $d_{34} = 0$ in the *c/a/c/a* state. The piezoelectric response $d_{35}$ increases with the misfit strain $S_m$ in films with the *c/a/c/a* and *ca*/aa*/ca*/aa** structures. In contrast, the coefficient $d_{34}$ displayed by the *ca*/aa*/ca*/aa** state decreases and even changes sign, as the magnitude of $S_m < 0$ is reduced for BT films. The nonmonotonic variation of $d_{34}$ = - $d_{35}$ in BT films having the $ca_1/ca_2/ca_1/ca_2$ structure is similar to the behavior of the dielectric constant $\varepsilon_{33}$ here (see Fig. 5c). Of course, the calculated piezoelectric responses $d_{3n}$ of the *c*/*a*/*c*/*a*, *ca**/*aa**/*ca**/*aa**, and $ca_1/ca_2/ca_1/ca_2$ states involve a nonzero domain-wall contribution.

If the relative domain population $\phi$ does not change during the piezoelectric measurements, Eqs. (31) and (32) yield the following relations for the coefficients $d_{34}$ and $d_{35}$:

$$d_{34} = Q_{44}\left[\phi\left(\eta'_{23}P'_3 + \eta'_{33}P'_2\right) + (1-\phi)\left(\eta''_{23}P''_3 + \eta''_{33}P''_2\right)\right], \qquad (35)$$

$$d_{35} = Q_{44}\left[\phi\left(\eta'_{13}P'_3 + \eta'_{33}P'_1\right) + (1-\phi)\left(\eta''_{13}P''_3 + \eta''_{33}P''_1\right)\right]. \qquad (36)$$

For the $a_1/a_2/a_1/a_2$ and $aa_1/aa_2/aa_1/aa_2$ polydomain states, where $P'_3 = P''_3 = 0$ at $\mathbf{E_0} = 0$ and $\eta'_{33} = \eta''_{33} = \eta_{33}$, from Eqs. (35)-(36) we obtain $d_{34} = Q_{44}\eta_{33}\langle P_2\rangle$ and $d_{35} = Q_{44}\eta_{33}\langle P_1\rangle$. These relations explain some peculiarities of the piezoelectric behavior revealed by numerical calculations (Figs. 6b and 7b). Indeed, they show that in ferroelectric films with the $a_1/a_2/a_1/a_2$ domain structure the equality $d_{34}$ = - $d_{35}$ must hold since $\langle P_1\rangle = -\langle P_2\rangle$ at the orientation of the $a_1/a_2$ walls shown in Fig. 1b. In the case of the $aa_1/aa_2/aa_1/aa_2$ polydomain state, the piezoelectric response $d_{34}$ is absent because the average polarization in the [010] direction equals zero ($\langle P_2\rangle = 0$) for the chosen domain geometry (Fig. 1c). The reduction of the piezoelectric constant $d_{35}$ with increasing misfit strain $S_m$ in films with the $a_1/a_2/a_1/a_2$ and $aa_1/aa_2/aa_1/aa_2$ structures can be attributed to the decrease of the dielectric susceptibility $\eta_{33}$ here (see Figs. 4c and 5c). The jump of $d_{35}$ at the transition from the $a_1/a_2/a_1/a_2$ to $aa_1/aa_2/aa_1/aa_2$ state in BT films is mainly due to a larger average polarization $\langle P_1\rangle$ in the latter.

## VI. CONCLUDING REMARKS

The nonlinear thermodynamic theory developed in this paper makes it possible to include laminar polydomain states into the misfit strain-temperature phase diagrams of epitaxial ferroelectric thin films and to determine the total dielectric and piezoelectric responses of polydomain films. Our calculations revealed several important features of the polydomain (twinned) states in epitaxial films of perovskite ferroelectrics. In particular, the phase diagram of BT films was found to be much more complicated than the diagram of PT films, which reflects the existence of three different ferroelectric phases (tetragonal, orthorhombic, and rhombohedral) in stress-free bulk BT crystals instead of only one (tetragonal) in PT crystals.

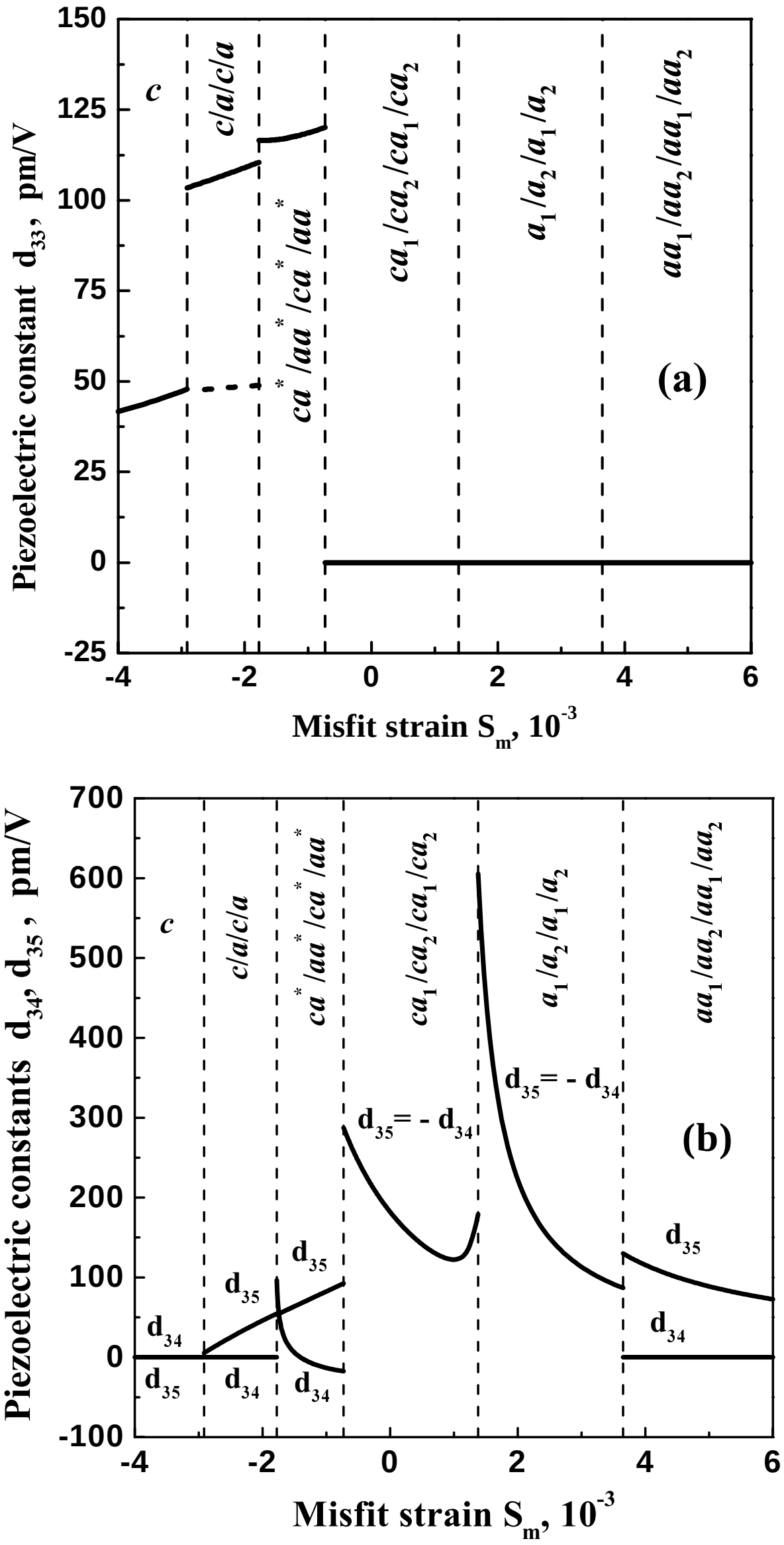

**FIG 7.** Piezoelectric coefficients of $BaTiO_3$ films as functions of the misfit strain $S_m$ at $T$ = 25°C: $d_{33}$ (a), $d_{34}$, $d_{35}$ (b). The dashed line shows the piezoelectric response $d_{33}$ of $BaTiO_3$ films with pinned domain walls calculated for the *c*/*a*/*c*/*a* state.

Surprisingly, the *direct transformation* of the paraelectric phase into the polydomain *c*/*a*/*c*/*a*

state was shown to take place in PT films only in a narrow misfit-strain range near $S_m = 0$ (Fig. 2a). Usually the formation of the *c*/*a*/*c*/*a* domain structure should proceed via a *two-stage process.*

At negative misfit strains, this process involves the appearance of the homogeneous ferroelectric *c* phase in the first stage followed by the nucleation of *a* domains inside this out-of-plane polarization state. In our approximation, the transformation of the *c* phase into the *c*/*a*/*c*/*a* pattern, which occurs along the thin line shown in Fig. 2a, appears to be continuous in the sense that the volume fraction $\phi_a = 1 - \phi_c$ of *a* domains gradually increases from zero on crossing the transition line $S_m^0(T)$. This is, however, not exactly correct because we have neglected the self-energies of domain walls in our calculations. Strictly speaking, the discussed transition is discontinuous, and a small but finite $\phi_a$ appears at the actual transition line, slightly shifted towards lower temperatures from the thin line drawn in Fig. 2a.

The nucleation of embedded *a* domains in the *c* phase during the cooling must be a thermally activated process so that a certain time is required for the formation of an equilibrium *c*/*a*/*c*/*a* pattern. Therefore, the observed populations of *c* and *a* domains in an epitaxial film may depend strongly on the cooling rate *dT*/*dt*. Slow cooling rates are expected to facilitate larger volume fractions $\phi_a$ to be observed at room temperature. In contrast, at very high rates *dT*/*dt* the complete suppression of domain formation may take place so that the metastable *c* phase becomes frozen in. Accordingly, our theory gives a natural explanation for the experimental results obtained in Refs. 5 and 14 on the effect of cooling rate on domain populations in epitaxial $PbZr_{0.20}Ti_{0.80}O_3$ and PT thin films. This explanation differs considerably from the one proposed earlier by Speck et al. on the basis of the linear domain theory.[7]

At positive misfit strains, the *c*/*a*/*c*/*a* structure is expected to form via the nucleation of *c* domains inside an in-plane polarization state already existing in the film. The equilibrium volume fraction of *c* domains, according to Eq. (16), must increase during the film cooling in this case. The critical temperature $T_{a\to c/a}$, at which the *c* domains begin to form in a film, may be well below the Curie-Weiss temperature $\theta = 479$°C of the bulk PT. Both these features were observed in PT films grown on (001)-oriented MgO by Lee and Baik, who used the synchrotron x-ray diffraction to measure the domain populations.[17] It should be emphasized that the critical temperature $T_{a\to c/a}$ may be much lower than the actual temperature of the ferroelectric phase transition $T_{aa}$, which is always higher than $\theta$.

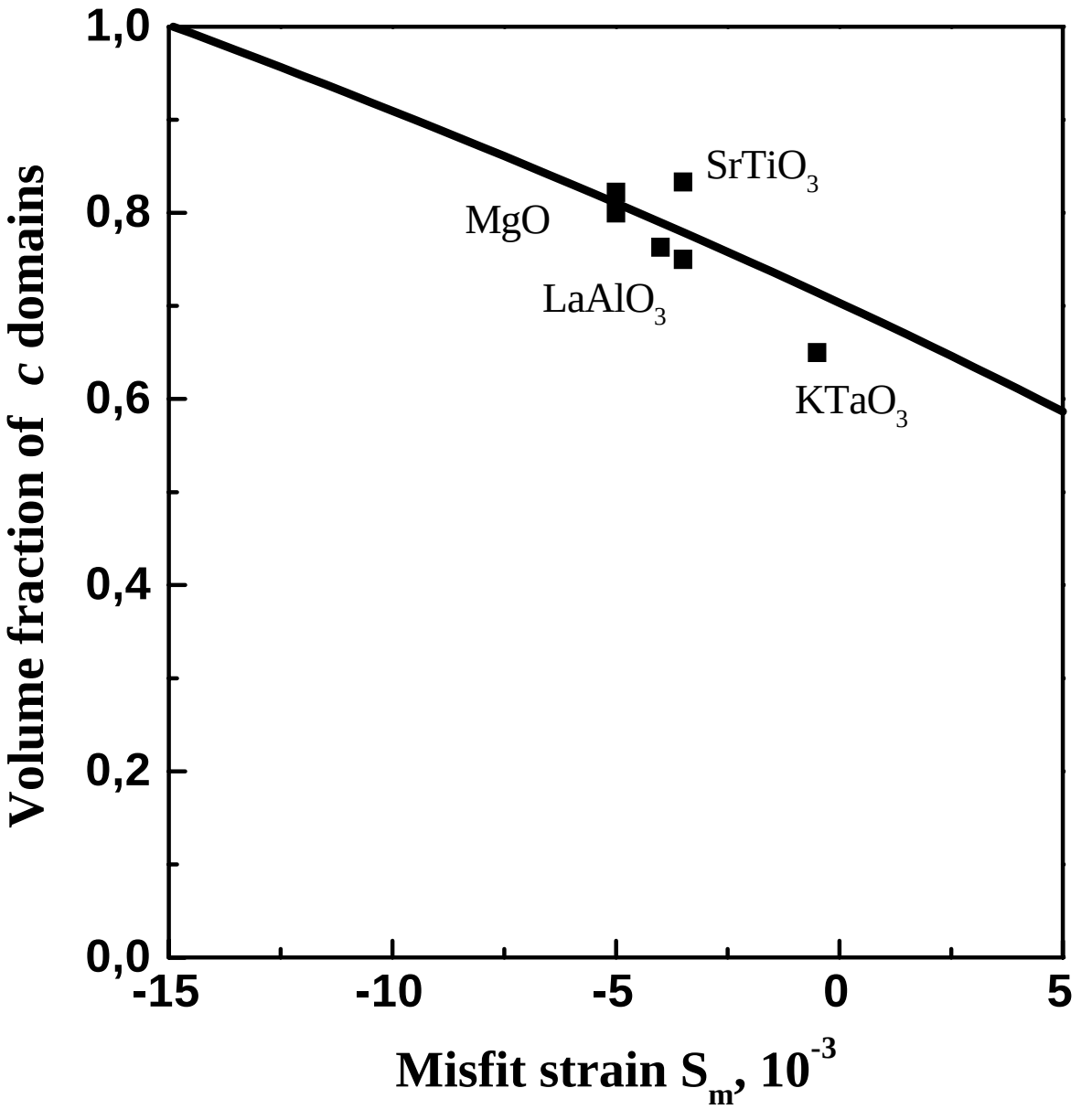

**FIG. 8**. Misfit-strain dependence of the volume fraction $\phi_c$ of *c* domains in thick $PbTiO_3$ epitaxial films at $T = 25$°C. The solid line gives the calculated fraction $\phi_c$. The squares show the values observed for films grown on various substrates indicated in the figure. The experimental data are taken from Refs. 11, 14 (MgO), 14 ($LaAlO_3$), 12, 14 ($SrTiO_3$), and 2 ($KTaO_3$). The misfit strains for corresponding heterostructures were calculated as described in Ref. 39 by using the following values of the thermal expansion coefficients: $5.5\times10^{-6}$ $K^{-1}$ (PT), $11\times10^{-6}$ $K^{-1}$ ($SrTiO_3$ and $LaAlO_3$),[53] $14.8\times10^{-6}$ $K^{-1}$ (MgO), $6.67\times10^{-6}$ $K^{-1}$ ($KTaO_3$).[6]

Finally, we would like to stress that our theoretical predictions on the geometry of the dense *c*/*a*/*c*/*a* structure agree with the experimental data available for sufficiently thick PT films ($H >> 100$ nm). Figure 8 compares the misfit-strain dependence of the volume fraction $\phi_c$ of *c* domains, calculated from Eqs. (15)-(16), and the measured values of $\phi_c$. Good agreement between the theory and experiment supports the validity of the stability range $R_d$, which has been predicted in this paper for the *c*/*a*/*c*/*a* polydomain state in the ($S_m$, $T$)-phase diagram. Unfortunately, we are not aware of experimental data on the dielectric and piezoelectric properties of polydomain single crystalline ferroelectric films. At the same time, the calculated dielectric constant $\varepsilon_{33}/\varepsilon_0 \approx 170$ and piezoelectric coefficient $d_{33} \approx 80$ pm/V of PT films near $S_m = 0$ are in order-of-magnitude agreement with the measured responses of polycrystalline PT films grown on platinum coated silicon substrates ($\varepsilon_{33}/\varepsilon_0 \approx 100$ and $d_{33} \approx 50$ pm/V).[54]

## ACKNOWLEDGMENT

The research described in this publication was made possible in part by Grant No. I/75965 from the Volkswagen-Stiftung, Germany.

---

*Electronic address: pertsev@domain.ioffe.rssi.ru